\documentclass[aps,prc,letterpaper,11pt,twoside,tightenlines,nofootinbib,showpacs,preprint]{revtex4}
\usepackage{graphicx}
\usepackage[sort&compress]{natbib}
\usepackage{subfigure}
\usepackage{amsmath}
\usepackage{amsfonts}
\usepackage{cancel}

\begin{document}

\arraycolsep1.5pt

\newcommand{\Ima}{\textrm{Im}}
\newcommand{\Rea}{\textrm{Re}}
\newcommand{\mev}{\textrm{ MeV}}
\newcommand{\be}{\begin{equation}}
\newcommand{\ee}{\end{equation}}
\newcommand{\ba}{\begin{eqnarray}}
\newcommand{\ea}{\end{eqnarray}}
\newcommand{\gev}{\textrm{ GeV}}
\newcommand{\nn}{{\nonumber}}


\title{
Asymmetry observables in $e^+ e^-\to \pi^+\pi^-\gamma$ in the $\phi$ region
within a chiral unitary approach  
}

\author{L.~Roca}
\affiliation{Departamento de F\'{\i}sica. Universidad de Murcia. E-30071, Murcia. Spain}

\author{E.~Oset}
\affiliation{
Departamento de F\'{\i}sica Te\'orica and IFIC, Centro Mixto Universidad de Valencia-CSIC,
Institutos de Investigaci\'on de Paterna, Aptdo. 22085, 46071 Valencia,
Spain}

\date{\today}

 \begin{abstract}

We make a theoretical study of the charge and 
forward-backward pion asymmetries in the
$e^+e^-\to\pi^+\pi^-\gamma$ process on and off the $\phi$ resonance
energy. These observables are rather sensitive to the inner details of the
theoretical models
to describe the reaction.
In addition to the standard implementation of the initial state
radiation (ISR) and the Bremsstrahlung contribution to the final state
radiation (FSR), we use the techniques of the chiral unitary approach to
evaluate the contribution from the mechanisms of $\phi$
 decay into   $\pi^+\pi^-\gamma$. This contribution involves the
 implementation of final state interaction from direct chiral loops, the
 exchange of vector and axial-vector resonances and the final state interaction through the consideration of the
 meson-meson unitarized amplitudes, which where found important in
a previous work describing the  $\phi\to\pi\pi\gamma$.
We find a good reproduction of the experimental data from KLOE for the
forward-backward asymmetry, both at the $\phi$ peak  and away from it. We 
also make predictions for the angular distributions of the charge asymmetry and
show that this observable is very sensitive to the chiral loops involved in
$\phi$ radiative decay.

\end{abstract}

\maketitle

\section{Introduction}
\label{Intro}

The radiative decays of the $\phi$ into $\pi \pi \gamma$ 
has been considered as one of the most suitable reactions
to get  information
about the $f_0(980)$ resonance
\cite{greco,franzini,colangelo,achasov,bramon5,Oller:2002na,lucio,uge,Markushin:2000fa,
Palomar:2003rb,Kalashnikova:2004ta,Bugg:2006sr,Escribano:2006mb,Escribano:2008xc}. 
In the last decade
the idea that 
 this resonance, as well as the other light scalar resonances, 
 $a_0(980)$, $f_0(600)$ or $\kappa(800)$, are dynamically generated 
 from multiple
scattering using the ordinary chiral
Lagrangians  \cite{npa,iam,nsd,Kaiser:1998fi} has shed new light into the problem of
the nature of the scalar mesons.
From the experimental point of view, both the CM2 collaboration at Novosibirsk 
\cite{Akhmetshin:1999di} and the KLOE collaboration of the 
$\phi$ factory DA$\Phi$NE at Frascati \cite{frascati1} reported results 
on the two pion invariant mass distribution in the 
$\phi\to\pi^0 \pi^0 \gamma$ decay.
The $\phi$ meson is produced from electron-positron collision.
In the study of the charged pion channel, $\phi\to\pi^+ \pi^- \gamma$,
 the problem of the large contribution of the initial state
radiation (ISR), (where the photon is emitted from the
electron or the positron, not possible in the $\pi^0 \pi^0 \gamma$
case for charge parity reasons), can turn itself into an advantage through the
analysis of different asymmetry observables, like the so called
forward-backward pion 
asymmetry  and charge asymmetry \cite{Binner:1999bt,Czyz:2002np}. 
The reason is
that this observable can be very sensitive to inner details of the
models to describe the reaction, thanks to the importance of the
interference between the final state radiation (FSR) and the ISR
mechanisms. The former is very model dependent and thus
 the study of this asymmetry is a good tool to test models for the FSR.
In particular, if the $e^+e^-$ center of mass
 energy is set to the $\phi(1020)$ peak, as is the case of DA$\Phi$NE,
 it is very suited to test $\phi$ decay models where the scalar
 mesons play a crucial role, particularly the $f_0(980)$ resonance.

For the FSR in the the $e^+ e^-\to \pi^+\pi^-\gamma$ reaction, there
are some standard models like the scalar QED (sQED) for the
final state Bremsstrahlung process
 \cite{Binner:1999bt,Czyz:2002np,Dubinsky:2004xv,Pancheri:2007xt}.
This is the most important contribution for large invariant mass of the
pions, but for the lowest part of the two pion spectrum other mechanisms related
to $\phi$ radiative decay become competitive. In 
 \cite{Dubinsky:2004xv,Pancheri:2007xt,Gallegos:2009qu} a correction 
from the vector contribution using the resonance chiral theory
Lagrangians \cite{egpdr} was also considered.
For the $\phi$ decay process there is a wider variety of
models which differ on the treatment of the scalar mesons.
Ref.~\cite{Czyz:2004nq}
considers the contribution of intermediate scalars 
using a point-like $\phi f_0\gamma$ interaction with explicit scalar
meson fields.
In Ref.~\cite{Pancheri:2007xt,Isidori:2006we}  
 the double resonance contribution
$e^+e^-\to\phi\to\rho^\pm\pi^\pm\to\pi^+\pi^-\gamma$ 
was also considered.
The most recent approach to the problem \cite{Gallegos:2009qu}
treats the scalar mesons from kaon loops using the techniques of the
chiral unitary approach to generate dynamically the scalar resonances
and compare the results with other models of scalar mesons like the
linear sigma model \cite{Bramon:2002iw}. The authors in \cite{Gallegos:2009qu}
could not find a good reproduction of the KLOE 
data \cite{Ambrosino:2005wk}  on the
asymmetry in the whole
double pion mass range\footnote{Recently the authors of 
ref.~\cite{Gallegos:2009qu} communicated us that their results in
the low energy region will be modified due to the numerics.}.
In ref.~\cite{Palomar:2003rb}, a very elaborate model was developed
 for the $\phi$ decay into $\pi\pi\gamma$. The model
considered the implementation of the two pseudoscalar final state
interaction using the techniques of the chiral unitary approach, both
from the kaon loops and from the production
through the exchange of intermediate vector and axial-vector
resonances. 
These new mechanisms were shown to be relevant 
for the two pion mass distribution in the $\phi$ radiative decay, specially at
the low part of the spectrum.

The aim of the present work is to evaluate the forward-backward  and
charge asymmetries in the $e^+ e^-\to \pi^+\pi^-\gamma$ reaction in order to
test the accuracy of the model used in ref.~\cite{Palomar:2003rb} to evaluate
the $\phi \pi\pi\gamma$ decay. This provides an extra test on the chiral unitary
approach and its repercussion on the dynamical generation of the light scalar
mesons.

\section{Formalism for $e^+ e^-\to \pi^+\pi^-\gamma$}
\label{sec:formalism}

The 
$e^+ e^-\to \pi^+\pi^-\gamma$ gets
contribution from two different processes depending on where the photon
is emitted: the
initial state radiation (ISR) and the final state radiation (FSR). In
the ISR the radiated photon is emitted from the initial
electron or positron and
it just involves a trivial electromagnetic process except for
the coupling of the pion
to the photon which can be accounted for by the
pion form factor.
In the FSR the final photon is emitted  after the virtual photon
attached to
the electron-positron line and it is the most model dependent part.
For a diagrammatic representation of the final diagrams for the ISR we
refer to fig.~1a), 1b) of ref.~\cite{Isidori:2006we}.

The amplitude for the FSR process can be decomposed in a model
independent way in terms of three
different structure functions.
For this decomposition we follow the formalism 
of ref.~\cite{Dubinsky:2004xv} which is
also used in ref.~\cite{Gallegos:2009qu}, 
(see ref.~~\cite{Dubinsky:2004xv} for further details).
We summarize it briefly in the present section.

For the $e^-(p_1) e^+(p_2)\to \pi^+(p_+)\pi^-(p_-)\gamma(k)$
it is convenient to introduce the variables
$Q=p_{1}+p_{2}$, $q=p_{+}+p_{-}$, $l=p_{+}-p_{-}$ and five 
independent Lorentz
scalars defined as
 \begin{align}
s  &  \equiv Q^{2}=2p_{1}\cdot p_{2},\nonumber\\
t_{1}  &  \equiv\left(  p_{1}-k\right)  ^{2}=-2p_{1}\cdot
k,\nonumber\\
t_{2}  &  \equiv\left(  p_{2}-k\right)  ^{2}=-2p_{2}\cdot k,\\
u_{1}  &  \equiv l\cdot p_{1}, \nn \\
u_{2}&    \equiv l\cdot p_{2}.\nonumber
\end{align}
(The electron mass is neglected in the present work).

As mentioned above, the total amplitude 
can be decomposed as

\be
T=T_{\textrm{ISR}}+T_{\textrm{FSR}}
\ee
with
\begin{align}
T_{ISR}  &  =-\frac{e}{q^{2}}L^{\mu\nu}\epsilon_{\nu}^{\ast}l_{\mu}
F_{\pi
}\left(  q^{2}\right)  ,\label{eq:TISR}\\
T_{FSR}  &  =\frac{e^{2}}{s}J_{\mu}T_{F}^{\mu\nu}\epsilon_{\nu
}^{\ast}, \label{eq:TFSR}
\end{align}
where
\begin{align}
L^{\mu\nu}  &  =e^{2}\overline{u}_{s_{2}}\left(  -p_{2}\right) \left[
\gamma^{\nu}\frac{\left(  -\cancel p_{2}+\cancel k+m_{e}\right)  }{t_{2}%
}\gamma^{\mu}+\gamma^{\mu}\frac{\left(  \cancel p_{1}-\cancel k+m_{e}\right)
}{t_{1}}\gamma^{\nu}\right]  u_{s_{1}}\left(  p_{1}\right)  ,\\
J_{\mu}  &  =e\overline{u}_{s_{2}}\left(  -p_{2}\right)  \gamma_{\mu}u_{s_{1}%
}\left(  p_{1}\right).
\end{align}
In Eq.~(\ref{eq:TISR}), $F_\pi(q^2)$ is the pion form factor which we take
in the present work from 
ref.~\cite{Dubinsky:2004xv}.

As explained in ref.~\cite{Pancheri:2007xt},
the most general 
form of the FSR tensor $T_{F}^{\mu\nu}$ can be written as 
\begin{equation}
T_{F}^{\mu\nu}=f_{1}\tau_{1}^{\mu\nu}+f_{2}\tau_{2}^{\mu\nu}+f_{3}\tau
_{3}^{\mu\nu}, \label{tensor}%
\end{equation}
where the $\tau_{i}^{\mu\nu}$ are
\begin{align}
\tau_{1}^{\mu\nu}  &  =k^{\mu}Q^{\nu}-g^{\mu\nu}k\cdot Q,\nonumber\\
\tau_{2}^{\mu\nu}  &  =k\cdot l\left(  l^{\mu}Q^{\nu}-g^{\mu\nu}k\cdot
l\right)  +l^{\nu}\left(  k^{\mu}k\cdot l-l^{\mu}k\cdot Q\right)  ,\\
\tau_{3}^{\mu\nu}  &  =Q^{2}\left(  g^{\mu\nu}k\cdot l-k^{\mu}l^{\nu}\right)
+Q^{\mu}\left(  l^{\nu}k\cdot Q-Q^{\nu}k\cdot l\right)  .\nonumber
\end{align}

The theoretical models to describe the FSR can then concentrate on
evaluating the Lorentz scalar functions, $f_i$.

The cross section for the $e^+ e^-\to \pi^+\pi^-\gamma$ reaction, with
normalization $\bar{u}_{s'} u_s=2 m_e\delta_{s,s'}$, can be
written as

\ba
\sigma=\frac{1}{16s(2\pi)^4}\int d\omega_+ \int d\omega_- 
\int d\cos\theta_+\int d\phi_{+-} |T|^2 \,\Theta(1-B^2)
\ea
where $\omega_{+(-)}$ is the energy of the $\pi_{+(-)}$, 
$\theta_+$ is the
$\pi^+$ polar angle, $\phi_{+-}$ is the azimuthal angle of the $\pi^-$
considering as $z$-axis the $\pi^+$ direction, $\Theta(x)$ is the step
function and $B$ is given by
\be
B=\frac{(\sqrt{s}-\omega_+-\omega_-)^2-|\vec p_+|^2-|\vec p_-|^2}
{2|\vec p_+||\vec p_-|}.
\ee  

The total amplitude squared, $|T|^2$, can be explicitly separated into
the contributions on the ISR, the FSR and the interference of the two
amplitudes:

\be
|T|^2=|T_{\textrm{ISR}}|^2+|T_{\textrm{FSR}}|^2
+2 \textrm{Re}\{T_{\textrm{ISR}}T^*_{\textrm{FSR}}\}
\label{eq:Tdecomp}
\ee
Due to charge parity ($C$) conservation, the final pion pair 
must be in a $C=-1(+1)$ state for the ISR(FSR) case.
The interference between two terms with opposite $C$-parity
is $C$-odd and, then, it changes sign under the interchange of the two
charged pions. Therefore, it produces a charge asymmetry, and also 
 a forward-backward asymmetry

\be
 A_{FB}=\frac{N(\theta_+> 90^{\circ})-N(\theta_+< 90^{\circ})}
{N(\theta_+> 90^{\circ})+N(\theta_+< 90^{\circ})},
\label{eq:FB}
\ee
where we consider $\theta_+$ defined with respect to the
positron beam, and $N$ represents the
number of $\pi^+$ events in the given angular region.

\section{UChPT model to the $\phi$ contribution to the FSR}
\label{sec:uchpt}

The expression of $|T_{\textrm{ISR}}|^2$, $|T_{\textrm{FSR}}|^2$ and 
$\textrm{Re}\{T_{\textrm{ISR}}T^*_{\textrm{FSR}}\}$ of
Eq.~(\ref{eq:Tdecomp}) in terms of the structure functions $f_i$ of 
Eq.~(\ref{tensor}) can be found in Eqs.~(8), (17) and (25) of
ref.~\cite{Dubinsky:2004xv}.
For the Bremsstrahlung process we use Eqs.~(11)-(20)
 of ref.~\cite{Pancheri:2007xt}, which correspond to the Feynman
 diagrams shown in fig.~2 of ref.~\cite{Dubinsky:2004xv}.

\begin{figure}[!t]
\begin{center}
\includegraphics[width=0.4\textwidth]{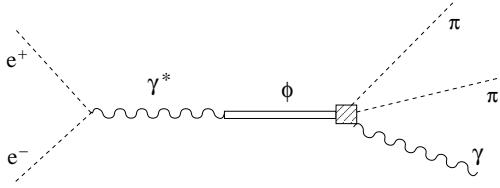}
\caption{Mechanism involving the $\phi(1020)$ decay.}
\label{fig:phicontrib}
\end{center}
\end{figure}

For $e^+e^-$ center of mass energies very close to the mass of the
$\phi(1020)$ resonance, the mechanisms producing a $\phi$ meson from
the virtual photon an its subsequent decay into $\pi^+\pi^-\gamma$ are
relevant, (see fig.~\ref{fig:phicontrib}).

\begin{figure}[!t]
\begin{center}
\includegraphics[width=0.7\textwidth]{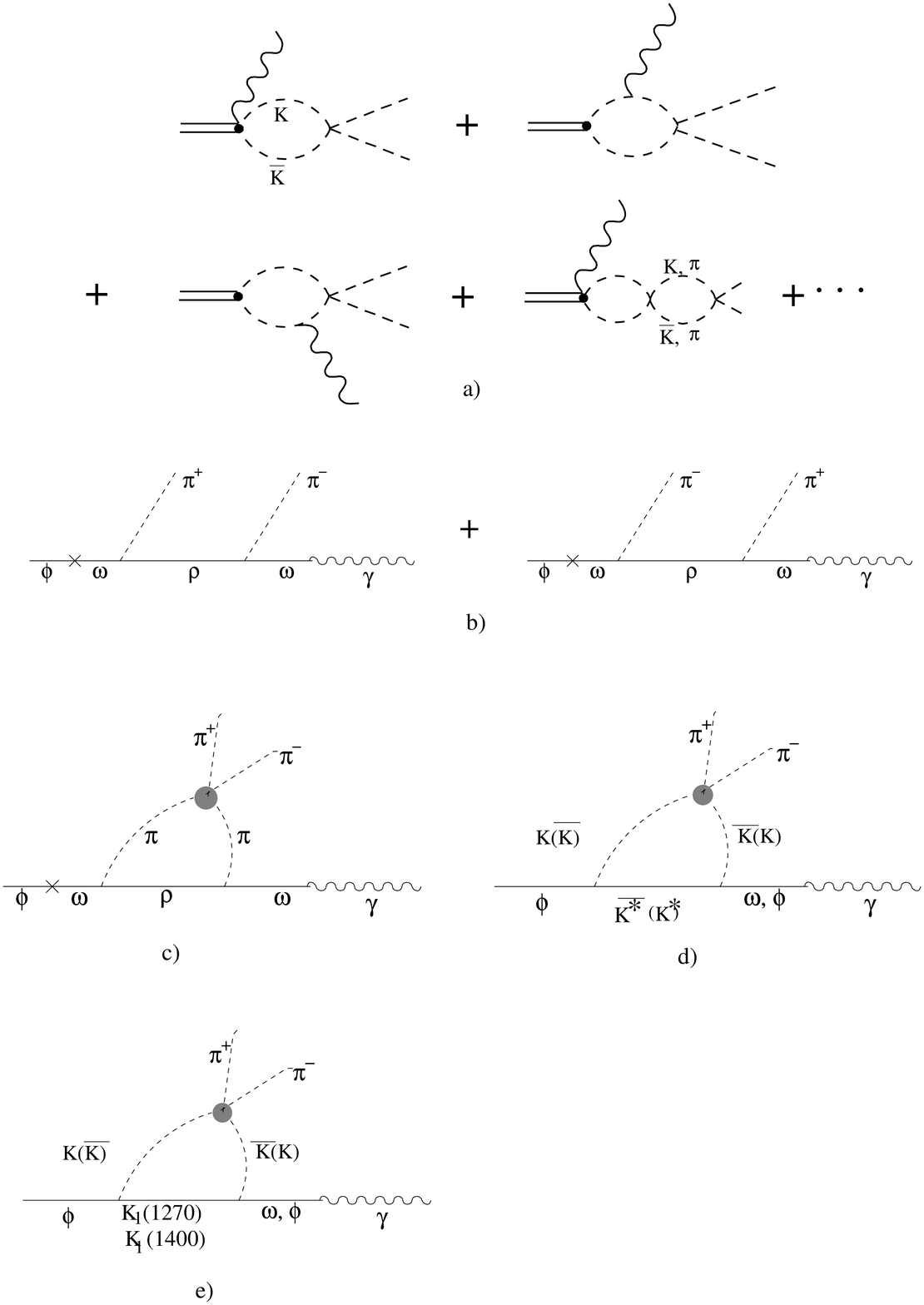}
\caption{Different contributions to the $\phi(1020)$ radiative 
decay in the model of \cite{Palomar:2003rb}.  a), chiral loops; 
b) vector meson exchange tree level; c)-e) loops of 
vector and axial-vector
exchange. The thick dots represent the unitarized 
meson meson scattering.
}
\label{fig:phi_mechs}
\end{center}
\end{figure}

The diagrams for the 
different contributions to the $\phi$ decay are  shown in
 fig.~\ref{fig:phi_mechs}.

This model contains, in a first place, the loops coming from $\phi\to
K^+ K^-$ decay and the implementation of the final state interaction of
the pions using the techniques of the chiral unitary approach
\cite{uge}, (fig.~\ref{fig:phi_mechs}a). This
mechanism was also considered in ref.~\cite{Gallegos:2009qu} using the
resonance chiral perturbation theory Lagrangians. In the present work we
use the Lagrangians of the hidden gauge symmetry \cite{hidden1,hidden2,hidden3},
where the conversion of the photons to vector mesons is a natural consequence of
the general Lagrangians. Their use is equivalent to working with the scheme of 
\cite{uge,Palomar:2003rb,Gallegos:2009qu} imposing the results of vector meson
dominance \cite{Ecker:1989yg}, essentially $F_V=2G_V$.

This mechanism contributes only to the $f_1$ function and is given by
\ba
f_1=-\frac{4\sqrt{2}}{3\sqrt{3}}D_\phi(Q^2) M_V^2
\widetilde{G}(\sqrt{s},M_I)
\frac{1}{Q^2-M_I^2}t_{KK,\pi\pi}^{I=0}(M_I)
\ea
where $M_I\equiv \sqrt{q^2}$ is the final two pion invariant mass, 
$\widetilde{G}(\sqrt{s},M_I)$ is the three meson loop function
given in \cite{Palomar:2003rb}, $D_\phi$ is the $\phi$ meson propagator
and $t_{KK,\pi\pi}^{I=0}$ is the s-wave isospin $I=0$ $K\bar K\to\pi\pi$ 
unitarized scattering amplitude in the normalization of \cite{npa}.
Note that the main difference with respect to \cite{Gallegos:2009qu} or
the expression given in \cite{uge,Palomar:2003rb} is the term
proportional to $F_V/2-G_V$ which is zero in vector meson dominance,
implicit in the HGS Lagrangians, and contributes very little
if explicitly considered \cite{uge,Gallegos:2009qu}.
It is worth mentioning that the meson-meson unitarized amplitude
generates also the
$\sigma(500)$ contribution apart from the $f_0(980)$ one, without the
need to include explicit fields for these scalar resonances. They appear
just from the implementation of unitarity form the lowest order
meson-meson chiral Lagrangian \cite{npa,iam,nsd}. Even more,
it provides the actual shape of the amplitude in the real axis (with
its possible background, etc),
 not just the pole
contribution.
The full gauge invariant set of diagrams in 
fig.~\ref{fig:phi_mechs}a requires also a term where the photon
couples to the four meson vertex. However, using the method of 
ref.~\cite{CloIsgKum} ones does not need to
evaluate its contribution explicitly
\cite{CloIsgKum,LucPer,Bra,Oller}.

The model~\cite{Palomar:2003rb} adds to the previous mechanisms 
the contribution from the
intermediate exchange of vector and axial-vector mesons,
 (fig.~\ref{fig:phi_mechs}b-e).
The vector meson exchange was also included, but only at tree level
(fig.~\ref{fig:phi_mechs}b),
 in 
\cite{Dubinsky:2004xv,Isidori:2006we,Pancheri:2007xt,Gallegos:2009qu}
in a different way. In \cite{Palomar:2003rb} the $\phi$ couples
to $\rho\pi$ through
$\phi \omega$ mixing since a direct coupling is OZI suppressed.
The exchange of axial-vector mesons was also included
 in \cite{Palomar:2003rb} but they are negligible
at tree level
and we also neglect them at tree level in the present work.

In our formalism, the contribution 
of the tree level vector meson
exchange  (fig.~\ref{fig:phi_mechs}b) to the structure functions  is given
by
\ba
f_1&=&\alpha[D_\rho(P_\rho) (l^2+Q\cdot k-2 k\cdot l)
          +D_\rho(P'_\rho)(l^2+Q\cdot k+2 k\cdot l)]\nn\\
f_2&=&-\alpha[D_\rho(P_\rho)+D_\rho(P'_\rho)]\nn\\
f_3&=&-\alpha[D_\rho(P_\rho)-D_\rho(P'_\rho)]
\label{eq:filoopchi}
\ea
with
\ba
\alpha=-{\cal C}\tilde{\epsilon}\frac{M_V^2}{9}
\frac{f^2G^2}{M_{\omega}^2}D_\phi(Q^2).
\label{eq:alpha}
\ea
See ref.~\cite{Palomar:2003rb} for further details on the
definition and values of the different constants of
Eq.~(\ref{eq:alpha}). In Eq.~(\ref{eq:filoopchi}) $P_\rho=(Q-l+k)/2$
and $P'_\rho=(Q+l+k)/2$.

One of the main novelties of the work \cite{Palomar:2003rb} was the
implementation of the final meson-meson scattering in the mechanisms
involving the vector and axial-vector exchange
 (figs.~\ref{fig:phi_mechs}c-e).
  These mechanisms modified significantly the
 shape of the double pion mass distribution in the $\phi$ 
 decay spectrum \cite{Palomar:2003rb}.
In addition to the loop mechanism constructed from the 
exchange of the $\rho$ meson (fig.~\ref{fig:phi_mechs}c),
it is also possible to
implement the loops in mechanisms with
exchange of vector $K^*$ (fig.~\ref{fig:phi_mechs}d) 
and axial-vector resonances, 
$K_1(1270)$ and $K_1(1400)$, (fig.~\ref{fig:phi_mechs}e).

For the evaluation of the asymmetry in the present work, 
these latter mechanisms (figs.~\ref{fig:phi_mechs}c-e)
contribute only to $f_1$ and the expressions can be obtained from the
amplitudes given in ref.~\cite{Palomar:2003rb}
substituting $\epsilon^*\cdot\epsilon$ by $(-M_V^2 \sqrt{2}
D_\phi(M_I))/(3eg  Q\cdot k)$. 
\\

\section{Results}
\label{sec:results}

In order to compare with the experimental data of
\cite{Ambrosino:2005wk,beltramethesis}, we implement in
our theoretical calculations
 the same
cuts than in the KLOE experiment. Thus, we account only for pions
 with angles
in the region $45^\circ<\theta_\pm<135^\circ$, and photons with 
$45^\circ<\theta_\gamma<135^\circ$ and $E_\gamma>10\textrm{ MeV}$.

  \begin{figure}[!t]
\begin{center}
\includegraphics[width=0.6\textwidth]{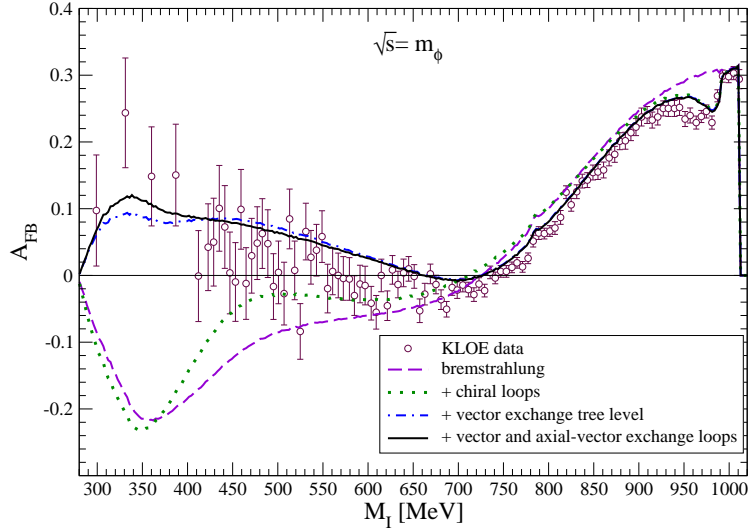}
\caption{Different FSR contributions to the forward-backward asymmetry
for electron-positron center of mass energy $\sqrt{s}=m_\phi$.
Experimental cuts are implemented.
Only Bremsstrahlung, dashed line; adding the direct chiral loops, dotted
line;  adding the
vector meson exchange at tree level, dashed-dotted line; full model,
solid line, which includes also the loops from the vector and
axial-vector exchange mechanisms. 
Experimental data from refs.~\cite{Ambrosino:2005wk,beltramethesis}. 
}
\label{fig:asymonpeak}
\end{center}
\end{figure}

 \begin{figure}[!t]
\begin{center}
\includegraphics[width=0.6\textwidth]{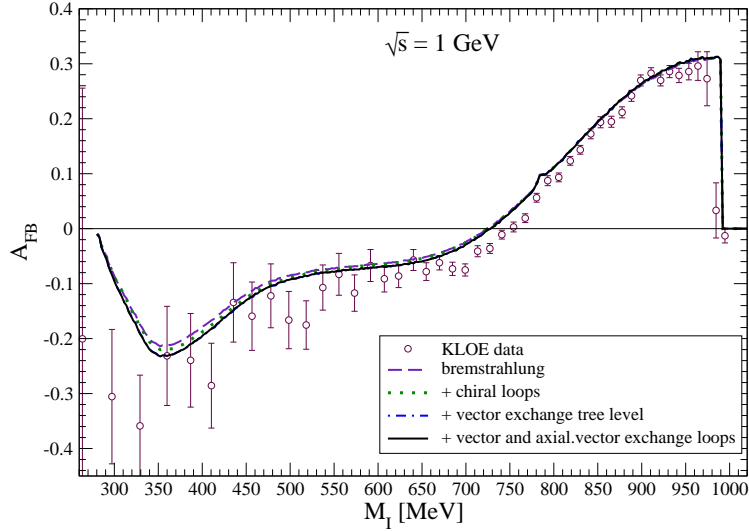}
\caption{Same as fig.~\ref{fig:asymonpeak} but for
 $\sqrt{s}=1\textrm{ GeV}$.
Experimental data from ref.~\cite{beltramethesis}}
\label{fig:asymoffpeak}
\end{center}
\end{figure}

In fig.~\ref{fig:asymonpeak}
we show the results for the forward-backward 
asymmetry as a function of the two pion
invariant mass, $M_I$, including only the Bremsstrahlung
process (dashed line); adding the direct
 chiral loops (dotted line); adding the
vector meson exchange at tree level (dashed-dotted line); and the full model 
(solid line) which includes also the loops from the vector and
axial-vector exchange mechanisms.

The sudden drop at the end of the spectrum is due to the photon
energy cut.
At higher invariant masses the dominant contribution is given by the
Bremsstrahlung process as already obtained in all previous theoretical
works on the reaction.
The effect of the scalar $f_0(980)$ resonance is clearly visible
 as a deep in the
invariant mass spectrum.
At low invariant masses the sequential
exchange of vector mesons has a crucial effect in the final shape of
the asymmetry.  This mechanism was not considered 
in ref.~\cite{Czyz:2004nq} claiming 
that is was negligible. On the other hand, the
implementation of the final loops in the sequential vector and
axial-vector meson exchange is less relevant than in the
$\phi\to\pi\pi\gamma$ decay \cite{Palomar:2003rb}.

If we change the center of mass energy of the electron-positron
collision to values off the $\phi$ peak,
 the contribution of the $\phi$ mechanisms is almost completely
removed. This is shown in fig.~\ref{fig:asymoffpeak},
where we plot the same
calculation as in fig.~\ref{fig:asymonpeak}
 but for $\sqrt{s}=1\gev$ instead of 
$\sqrt{s}=m_\phi=1.02\gev$ used in fig.~\ref{fig:asymonpeak}. 
The experimental data are taken from ref.~\cite{beltramethesis}. We can see 
that the standard, non $\phi$, mechanisms suffice
 to obtain a good agreement with the KLOE data \cite{beltramethesis}.
In order to cancel the $\phi$ effects it is not necessary to move
very much the energy from  
the $\phi$ peak since the $\phi$ resonance is very narrow,
$\sim 4\mev$.   
Clearly now the effect of the $f_0(980)$ is negligible,
since it is part of  the
$\phi$ production.

Overall, we find a good reproduction of this asymmetry in the whole
invariant mass spectrum.

\begin{figure*}
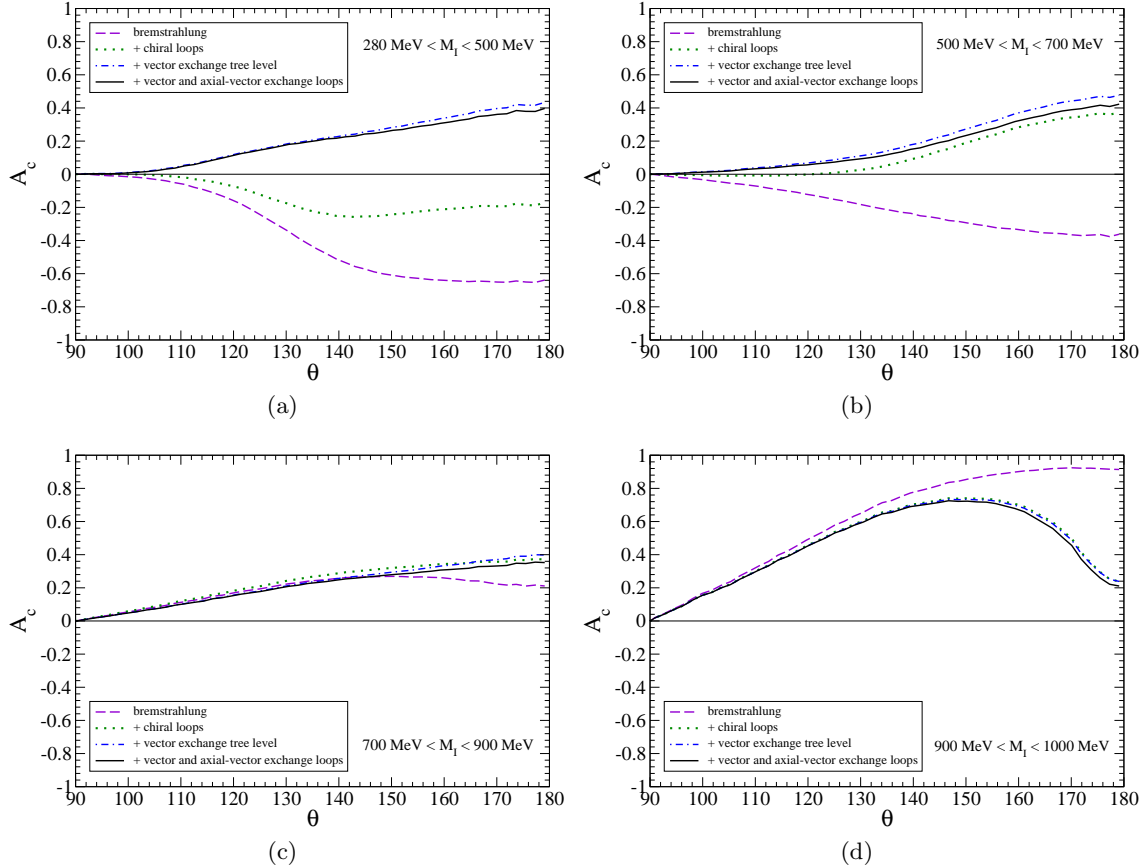

     \centering
     \subfigure[]{
          \label{fig:chargeasym1}
          \includegraphics[width=.45\linewidth]{figure5a.eps}}
     \subfigure[]{
          \label{fig:chargeasym2}
          \includegraphics[width=.45\linewidth]{figure5b.eps}}
\\
     \subfigure[]{
          \label{fig:chargeasym3}
          \includegraphics[width=.45\linewidth]{figure5c.eps}}
    \subfigure[]{
          \label{fig:chargeasym4}
          \includegraphics[width=.45\linewidth]{figure5d.eps}}
     \caption{Different FSR contributions to the charge asymmetry,
     eq.~\ref{eq:Ac},
for electron-positron center of mass energy $\sqrt{s}=m_\phi$.
Experimental cuts for the photons are implemented.}
     \label{fig:chargeasym}
\end{figure*}

We can use other observables to test the theoretical model and the
contribution of the different mechanisms. For
instance, like in ref.~\cite{Czyz:2004nq}, we can calculate the charge
asymmetry
\be
A_c(\theta)=\frac{N_+(\theta)-N_-(\theta)}{N_+(\theta)+N_-(\theta)}
\label{eq:Ac}
\ee
\noindent
where $N_{+(-)}(\theta)$ is the number of $\pi^{+(-)}$ emitted in the
$\theta$ direction defined with respect the positron axis. As discussed after
eq. (\ref{eq:Tdecomp}), only the interference term
$\textrm{Re}\{T_{\textrm{ISR}}T^*_{\textrm{FSR}}\}$ changes sign upon the
interchange of $\pi^+$, $\pi^-$, hence the numerator of eq. (\ref{eq:Ac}) contains
only this term in $|T|^2$.
 This asymmetry satisfies
$A_c(\theta)=-A_c(180^\circ-\theta)$, thus we only plot angles
 from
$90^\circ$ on.
We plot in
fig.~\ref{fig:chargeasym} the charge asymmetry
as a function of the polar angle of the corresponding pion, for
different $\pi\pi$ invariant mass ranges and implementing
the KLOE acceptance for the photons.
There is no experimental data published on this observable.
Again one can see in the figures the strong effect of the vector
meson exchange mechanisms at low invariant masses and the $f_0(980)$
at high masses.

The effect of the chiral loops in $A_c$ is very
important in all ranges of the invariant mass. This effect was not so pronounced for
the forward-backward asymmetry. In the range of 
$500\mev\le M_I \le 700\mev$
the chiral loops reverse the sign of the $A_c$ 
magnitude, something that could be
clearly visible with the present KLOE angular acceptance.  In the range of
 $900\mev \le M_I \le 1000\mev$ the chiral loops reduce considerably the
 strength of $A_c$, particularly close to $180^\circ$. 
   In \cite{Czyz:2002np} it was shown that the amplitude $A_c$ is very sensitive
   to details of the model for $\phi$ radiative decay. In this respect, it is
   important to note that when applying the chiral unitary 
   approach to the present problem we have no
   freedom in parameters and the results presented here are a neat prediction of
   the model. Since the chiral unitary approach used is the one that generates
   dynamically the $f_0(980)$ and $f_0(600)$, an eventual agreement
   of the experiment with the predictions would provide 
   extra support for this
   interpretation of the nature of these resonances.

\section{Conclusions}
\label{sec:concl}

We have calculated the different contributions to  
the  forward-backward and charge pion asymmetries in  
$e^+ e^-\to \pi^+\pi^-\gamma$. The main aim has been to test
the chiral unitary approach calculation of the $\phi\to\pi\pi\gamma$
decay \cite{Palomar:2003rb}. 
This model implements the final state interaction from direct $\phi\to
K\bar K$ decay, the sequential exchange of vector and axial-vector
resonances at tree level and the final state
interaction of the meson pair. 
This meson-meson rescattering generates the scalar resonances without
the need to include them as an explicit degree of freedom. 

The results of the present work show that there is a good agreement
of our theoretical model with the experimental KLOE data on the
forward-backward asymmetry, both on the $\phi$ peak as well as
outside its range. The test done outside the $\phi$ peak using the
data of \cite{beltramethesis} indicates that one has a good control
on the conventional non $\phi$ mechanisms of initial and final
state radiation  in $e^+e^-\to \pi^+\pi^-\gamma$. The changes seen
in the asymmetry at the $\phi$ peak can then clearly be attributed
to the $\phi$ radiative decay mechanisms. Yet, theses changes,
particularly at low $\pi\pi$ invariant masses where they are more
drastic, are mostly due to the sequential vector exchange mechanism
at tree level, although in the intermediate range of invariant
masses the chiral loops are relevant. From the purpose of finding
observables which are very sensitive to these chiral loops, we
found even more interesting the charge asymmetry. There we could
see that at low invariant masses the chiral loops are important,
and in the intermediate range $500\mev<M_I<700\mev$, they even
change the sign of the observable. At higher invariant masses the
effects are also remarkable, particularly at angles close to 
$180^\circ$, outside the present range of KLOE.

The present results should encourage experimental efforts to
measure the charge asymmetry and other related observables which
could shed more light on the nature of the scalar resonances.

\section*{Acknowledgments}

We thank M.~Napsuciale and A.~Gallegos for helpful discussions.
This work is partly supported by DGICYT contracts  FIS2006-03438,
FPA2007-62777 and 
the EU Integrated Infrastructure Initiative Hadron Physics
Project  under Grant Agreement n.227431.

\end{document}